\documentclass[prl,twocolumn,aps,showpacs,superscriptaddress,noshowpacs]{revtex4}
\usepackage{bm}
\usepackage{graphicx}
\usepackage{amssymb}

\begin{document}

\newcommand\kB{k_\mathrm{B}}
\newcommand\Rw{R_\mathrm{Wall}}
\newcommand\Energy{\mathcal E}

\newcommand\muHbar{\mu_{\bar \mathrm{H}}}
\newcommand\muHbarb{\bm{\mu}_{\bar \mathrm{H}}}

\newcommand\Mg{M_\mathrm{g}}
\newcommand\Q{Q_{\bar \mathrm{H}}}
\newcommand\pbar{\bar \mathrm{p}}

\newcommand\rhat{{\hat \mathbf{r}}}
\newcommand\thetahat{{\hat{\bm\theta}}}
\newcommand\zhat{{\hat\mathbf{z}}}
\newcommand\xhat{{\hat\mathbf{x}}}
\newcommand\yhat{{\hat\mathbf{y}}}

\newcommand\avzR{\langle z\rangle_\mathrm{R}}
\newcommand\avzL{\langle z\rangle_\mathrm{L}}
\newcommand\avz{\langle z\rangle}
\newcommand\avzD{\langle z\rangle_\Delta}

\newcommand\ER{E_\mathrm{R}}
\newcommand\EL{E_\mathrm{L}}

\newcommand\Deltapot{\langle\Phi\rangle_\Delta}
\newcommand\DeltaEk{\Delta E_{\mathrm{K}}}
\newcommand\Depth{U_{\mathrm{T}}}

\newcommand\QQ{-1.3}
\newcommand\samperr{1.1}
\newcommand\samperrninety{1.8}
\newcommand\syserr{0.4}
\newcommand\syserrninety{0.6}

\newcommand\THeat{t_{\rm{a}}}
\newcommand\thalf{t_{1/2}}
\newcommand\DeltaPhi{\Phi_\Delta}
\newcommand\tstep{\bar{t}}
\newcommand\tsigma{\sigma_{\tstep}}

\newcommand\pos{\mathbf{r}}

\title{Using stochastic acceleration to place experimental limits on the charge of antihydrogen}
\author{M. Baquero-Ruiz}
\author{A. E. Charman}
\affiliation{Department of Physics, University of California at Berkeley, Berkeley, CA 94720-7300, USA}
\author{J. Fajans}
\affiliation{Department of Physics, University of California at Berkeley, Berkeley, CA 94720-7300, USA}
\affiliation{Lawrence Berkeley National Laboratory, Berkeley, California 94720, USA}
\author{A. Little}
\author{A. Povilus}
\affiliation{Department of Physics, University of California at Berkeley, Berkeley, CA 94720-7300, USA}
\author{F. Robicheaux}
\affiliation{Department of Physics, Purdue University, West Lafayette, Indiana 47907, USA}
\author{J. S. Wurtele}
\affiliation{Department of Physics, University of California at Berkeley, Berkeley, CA 94720-7300, USA}
\affiliation{Lawrence Berkeley National Laboratory, Berkeley, California 94720, USA}
\author{A. I. Zhmoginov}
\affiliation{Department of Physics, University of California at Berkeley, Berkeley, CA 94720-7300, USA}

\date{Received \today}

\begin{abstract} Assuming hydrogen is charge neutral, CPT invariance demands that antihydrogen also be charge neutral. Quantum anomaly cancellation also demands that antihydrogen be charge neutral.  Standard techniques based on measurements of macroscopic quantities of atoms cannot be used to measure the charge of antihydrogen.  In this paper, we describe how the application of randomly oscillating electric fields to a sample of trapped antihydrogen atoms, a form of stochastic acceleration, can be used to place experimental limits on this charge.
\end{abstract}


\maketitle

\section{Introduction}
In 2010, antihydrogen atoms were trapped at CERN \cite{andr:10a,andr:11a,andr:11b,gabr:12}. Since then, CERN's ALPHA collaboration has reported initial experimental results on the two most commonly discussed symmetry tests with antihydrogen: spectral tests of CPT \cite{amol:12a}; and gravity freefall tests of the weak equivalence principle \cite{amol:13}.  Future experiments are expected to obtain much more precise results \cite{kell:08,char:11a,ashk:12,amol:13,zhmo:13,cesa:09a,hami:14}.

A less commonly discussed test of fundamental symmetries concerns the electric charge of antihydrogen.  Assuming that atomic hydrogen (H) is itself charge neutral, CPT demands that antihydrogen also be charge neutral.  While there do not appear to be extraordinarily precise measurements on H itself, other normal-matter atoms and molecules are known to be neutral to remarkable precision \cite{bres:11}: to about $10^{-21}e$ for diverse species such as He, H$_2$, and SF$_6$, where $e$ is the elementary charge.  The methods used in these studies are unsuitable for antihydrogen as they require macroscopic quantities of atoms or molecules; to date, only about $500$ antihydrogen atoms have been trapped and detected, and there are no prospects for trapping macroscopic quantities.   Charge neutrality of antimatter atoms, as well as of matter atoms, is also expected from the condition for quantum anomaly cancellation, which is required for theoretical consistency in quantum field theory \cite{quig:97}.

ALPHA recently reported \cite{amol:14a} a bound on the antihydrogen charge $Qe$ of $Q=(\QQ\pm\samperr\pm\syserr)\times 10^{-8}$, where the first error arises from counting statistics, and the second error is estimated based on systematic effects. This bound was based on a search for the deflection of putatively charged antihydrogen atoms by an applied electric field.  Here we describe a different technique, related to stochastic acceleration \cite{stur:66,tsus:09}, to measure the charge.  This technique uses randomly time-varying electric fields to eject putatively charged antihydrogen atoms from an ALPHA-style trap.  Current measurements using this technique \cite{amol:14a} set a bound on $Q$ of about $2\times 10^{-7}$, an order of magnitude looser than that found by the deflection technique, but this stochastic acceleration technique can easily be extended to much better precision, perhaps ultimately reaching $10^{-12}$.


\section{Apparatus}
\begin{figure}[tb!]
\centerline{\includegraphics[width=3.35in]{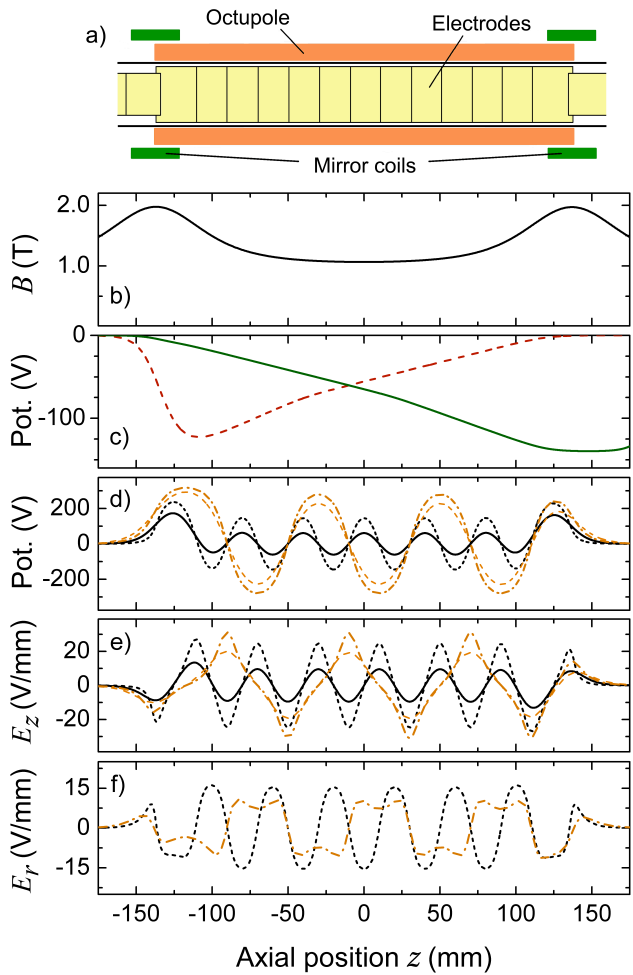}}
\caption{{\bf Experimental summary} (a) A schematic of the antihydrogen production and trapping region of the ALPHA apparatus, showing the cryogenically cooled cylindrical Penning-Malmberg trap electrodes, and the mirror and octupole magnet coils. The ALPHA positron source (not shown) is towards the right, and the Antiproton Decelerator (not shown) is towards the left.  (b) The on-axis magnetic field $B$ as a function of $z$.  (c) Typical on-axis electrostatic potentials used in the prior \cite{amol:14a} deflection-style experiments. (d) Values of the electrostatic potential at $r=0$ (black solid line) and at $r=0.6\Rw=13.6\,$mm (black dashed line) for a possible stochastic acceleration measurement.  Here, biases alternating between $\pm 350\,$V are applied to consecutive electrodes in the region between the magnetic field maxima; all other electrodes are kept at $0\,$V. Also shown are the $r=0$ and $r=0.6\Rw$ potentials when pairs of contiguous electrodes are joined and alternated at $\pm 350\,$V (orange dashed line and orange dashed-dotted line). (e) Axial component of the electric field for the potentials in (d). (f) Radial component of the electric fields for the potentials in (d), evaluated at $r=0.6\Rw$. Graphs (e) and (f) use the same line identification scheme as (d).}
\label{Fig1SchematicFields}
\end{figure}

ALPHA traps antihydrogen atoms by producing and capturing them in a minimum-B trap \cite{prit:83}. The trap confines those anti-atoms whose magnetic moment $\muHbarb$ is aligned such that they are attracted to the minimum of the trap magnetic field $\mathbf{B}$, and whose kinetic energy is below the trap well depth, $\muHbar(|\mathbf{B}|_\mathrm{Wall}-|\mathbf{B}|_\mathrm{Center})$. In ALPHA (see Fig.~\ref{Fig1SchematicFields}a), this magnetic minimum is created by an octupole magnet which produces transverse fields of magnitude $1.54\,$T at the trap wall ($\Rw=22.3\,$mm), and two mirror coils which produce axial fields of $1\,$T at their centers.  The mirror coil centers are symmetrically located at distances $z=\pm 137\,$mm from the trap center at $z=0$ (see Fig.~\ref{Fig1SchematicFields}b).  These fields are superimposed on a uniform axial field of 1T produced by an external solenoid \cite{bert:06,andr:08b}; taken together, they create a trap of depth $540\,$mK \cite{amol:14a}.

The general methods by which anti-atoms were produced from antiprotons and positrons, and then captured in the ALPHA trap, are described in Refs.~\citenum{andr:10a,andr:11a,andr:11b,andr:11d}; in this article we concentrate only on the last two phases of the ALPHA experiment, where anti-atoms were first held in static magnetic fields for times up to $1000\,$s, and then released from the minimum-B trap by gradually turning off the octupole and mirror fields.  The escaping anti-atoms were then detected with about $60$\% efficiency \cite{andr:12} when they annihilated on the trap wall.

\section{Charged Particle Deflection Experiments}
In the previously reported deflection experiments \cite{amol:14a}, the anti-atoms were subjected to electric fields similar to those derived from the potential shown in Fig.~\ref{Fig1SchematicFields}c. Together with the magnetic field, these electric fields form a well in which the on-axis potential energy of an anti-atom with a putative charge $Qe$ is given by
\begin{equation}
U(z)=\muHbar B(z)-\frac{QeE}{k_b}z,
\label{potential}
\end{equation}
where all energies are specified in units of kelvins, $\muHbar=0.67\,$K/T is the normalized antihydrogen magnetic moment \cite{amol:12a}, $k_b$ is the Boltzmann constant, and where we approximate the electric field inside the trap by a constant value $E$.  As $B(z)$ has a minimum at $z=0$, this well also has a minimum at $z=0$ when $Q=0$.  Consequently, the annihilations that result from the last-phase magnet shutoff will be centered around $z=0$.  If $Q\neq 0$, then the well minimum will shift \cite{amol:14a} by an amount
\begin{equation}
\avzD \propto QE/\beta,
\label{DeflectionScaling}
\end{equation}
where we approximate the magnetic field around the minimum as $B(z)=B_0+\beta z^2$.  To set the deflection-based $Q$ bound quoted above \cite{amol:14a}, ALPHA used measurements of the experimental $\avzD$, coupled with extensive computer simulations to determine Eq.~(\ref{DeflectionScaling})'s scaling constant.

The scaling in Eq.~(\ref{DeflectionScaling}) suggests three methods of tightening the bounds on $Q$: (i) Increasing $E$.  Unfortunately, arcing and other experimental concerns limit any increase in $E$ to factors in the range of 2 to 4. (ii) Decreasing $\beta$.  The increase in $B(z)$ going from the trap center to the trap axial ends, which is proportional to $\beta$, sets the trap depth. Currently, the trap depth of $540\,$mK cannot be lowered without anti-atoms escaping because many of the anti-atoms are only shallowly trapped \cite{andr:11a,amol:12}.  Laser cooling of the trapped anti-atoms \cite{donn:13} has the potential to lower the anti-atom temperature to about $20\,$mK, which would allow us to lower the post-cooling trap well depth to perhaps $30\,$mK without losing too many anti-atoms; $\beta$ would decrease correspondingly. (iii) Obtaining a better experimental determination of the measured $\avzD$.  The error in $\avzD$ is set by counting statistics.  ALPHA utilized a sample of $386$ anti-atoms collected over two years for its determination \cite{amol:14a} of $Q$. As the statistical error scales as the inverse square root of the number of samples, it would be difficult to decrease this error without significantly increasing the trapping rate.  Taken together, these improvements have the potential to tighten the bound on $Q$  by less than a factor of one hundred.  Consequently, it is worth investigating other methods to determine $Q$.

\section{Stochastic Acceleration Methodology}
\subsection{Charged Particle Ejection}
The electric fields $E$ effectively remove any inadvertently mirror-trapped antiprotons from the system (cf.\ Table~1 of Ref.~\citenum{amol:12}). However, other particles, such as putatively charged antihydrogen atoms, can also be ejected, and at a charge much less than the unit charge.  For the potential in the Fig.~\ref{Fig1SchematicFields}c, the wells predicted by Eq.~(\ref{potential}) will cease to exist for any anti-atom with a $|Q|\gtrsim 2\times 10^{-6}$.  Further, anti-atoms with a charge below that required for the well to disappear, i.e. $0<|Q|<2\times 10^{-6}$, will still be accelerated by the application of the electric fields.  They can be ejected if the extra increment of energy they gain is sufficient for them to climb over the trap walls.

\begin{figure}[h]
\centerline{\includegraphics[width=3.35in]{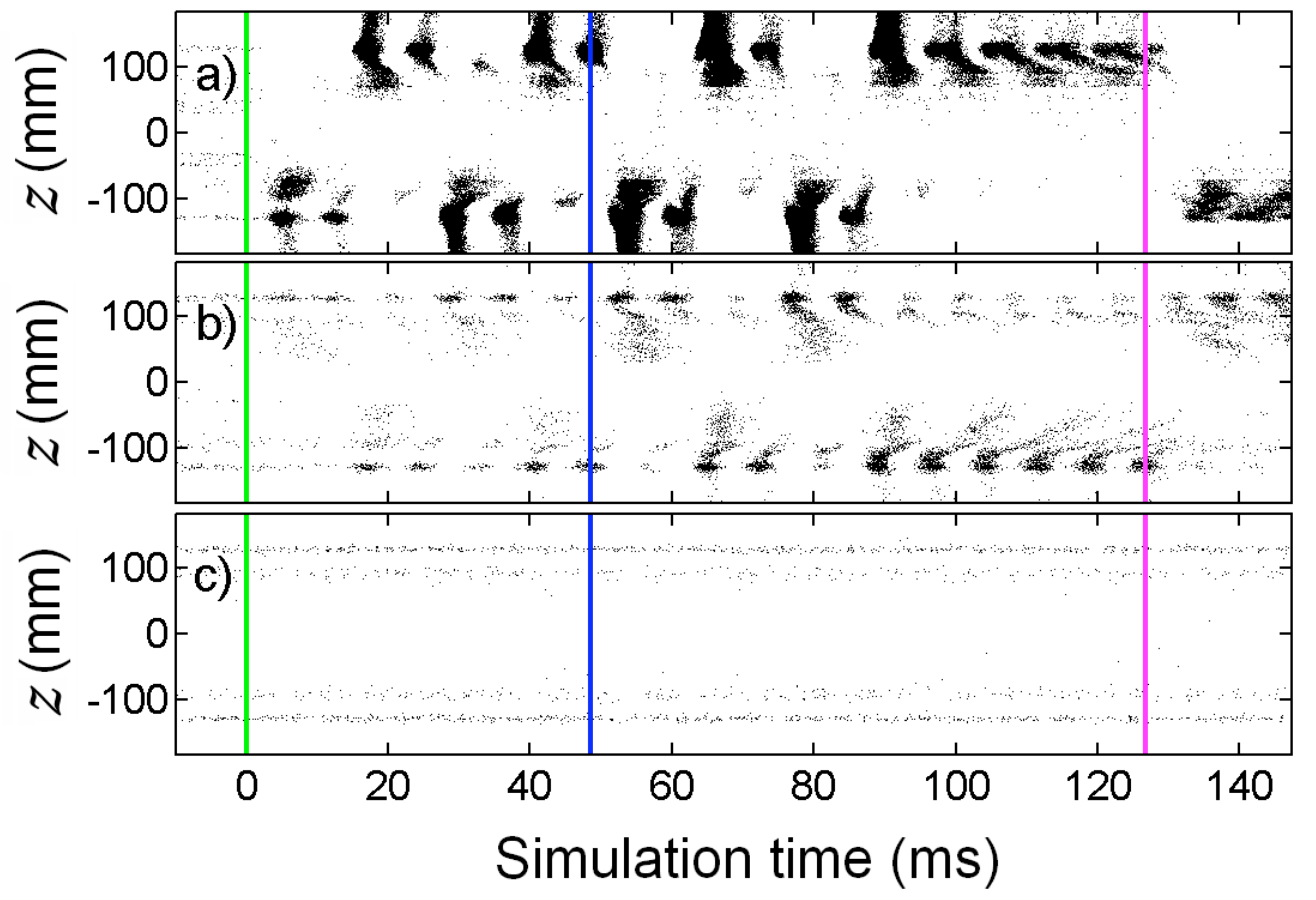}}
\caption{{\bf Clearing simulation results} Simulated axial annihilation locations $z$ versus annihilation time for different values of $Q$: (a) $Q=2\times 10^{-7}$, (b) $Q=-3\times 10^{-8}$, and (c)  $Q=0$. Each dot represents one annihilation.  The colored vertical lines indicate the start time for different manipulations: the half-strength clearing cycles begin at the leftmost green line, the full-strength clearings cycles begin at the middle blue line, and the final $E$ field (used in the deflection measurement) is applied at the magenta rightmost line.  The multiple clumps in (a) and (b) reflect multiple oscillations of the anti-atoms in the potential well of Eq.~(\ref{potential}).  The very small losses in (c) are due to the gradual depopulation of the quasi-trapped states \cite{coak:05,andr:10a,amol:12}. In (a), 54\% of the charge is lost from $t=0$ to the end of the plot; in (b), 2.4\% of the charge is lost; in (c), 0.3\% of the charge is lost.}
\label{EjectionSim}
\end{figure}

In the previously reported experiment \cite{amol:14a}, the fields were not static; they cycled between potentials similar to the two shown in Fig.~\ref{Fig1SchematicFields}c.  Altogether, the fields underwent nine transitions. (The first four transitions used half-strength fields).  The transition timescales ($\tstep\approx 2\,$ms) were short compared to the anti-atom orbit timescales (typically $\lesssim 10\,$ms), and the orbit timescales were comparable to the time intervals between transitions ($12\,$ms).  Thus, the accelerations at each transition were non-adiabatic, and, at each transition, the anti-atoms received a kinetic energy ``kick.''  These kicks were effectively random; an individual kick might have increased or decreased the anti-atom's energy. Depending on $Q$ and the vagaries of the stochastic process, the kicks might give the anti-atoms enough energy to escape the well.  This process is illustrated by the simulation results in Fig.~\ref{EjectionSim}.  Large numbers of anti-atoms were ejected for large $Q$, small numbers for small $Q$, and very few anti-atoms were lost when $Q=0$.  As approximately half of the simulated anti-atoms were lost at $Q=2\times 10^{-7}$, but experimentally ALPHA observed trapped anti-atoms at the end of these cycles, ALPHA set an experimental limit in the neighborhood of  $|Q|<2\times 10^{-7}$ for the anti-atom charge \cite{amol:14a}.  However, the absence of an absolute trapping rate measurement makes setting a precise limit impossible for this current dataset using this methodology.

\subsection{Experimental Design}
We propose \cite{amol:14a,baqu:13} to remedy this problem, and make an improved determination of $Q$, by measuring the trapping rate with and without the application of stochastic electric fields. Specifically, we would measure the number of anti-atoms remaining in the trap (by turning off the trapping magnetic fields) per trapping attempt, or trial, after a set of acceleration cycles were applied for a time $\THeat$, or else after the trap was kept quiescent (no accelerating fields) for the same time $\THeat$. By ensuring that both types of trials hold the anti-atoms for equal times, we would take into account any effects from vacuum annihilation or other anti-atom loss mechanisms.  To ameliorate the effects of long time drifts, we would alternate the two types of trials.  If we were to observe that the acceleration-on rate was lower than the acceleration-off rate, we could use analytic or numerical means to estimate the $Q$ that would cause the observed difference.  If, as is more likely, we observe no statistically significant difference between these rates, we could use these same means to calculate the $Q$ that would have caused a measurable difference; a good criterion would be to find the critical $Q$ which would cause half the anti-atoms to be lost.  In either case, this stochastic acceleration methodology would place a value or bound on $Q$.

\subsection{Approximate Determination of $Q$}
Let us take $\pm\Deltapot$ to be the typical electrical potential change experienced by an anti-atom during one kick. Such a potential change would result in a center-of-mass kinetic energy change of $\DeltaEk \sim \pm Qe\Deltapot$ for an anti-atom with a charge $Q$.  Following the classic random walk scaling, the typical total energy gained after $N$ kicks would be on the order of $|Q|e\Deltapot\sqrt{N}$.  If this energy exceeds the trap depth $\Depth$, a condition approximately met if
 \begin{equation}
|Q| \gtrsim \frac{\Depth}{e\Deltapot}\sqrt{\frac{1}{N}},
\label{StochasticLimit}
\end{equation}
then the kicks will drive a typical anti-atom of charge $Q$ out of the trap.

\subsection{Advantages of Stochastic Acceleration}
This stochastic acceleration methodology has several advantages over the deflection methodology employed in Refs.~\citenum{amol:14a,baqu:13} and could ultimately lead to a much better bound. First, instead of searching for a small average deflection, which requires hundreds of anti-atoms, the stochastic acceleration methodology relies on the simpler observation that anti-atoms have survived the acceleration cycles; thus, the test can reach statistical significance with only a few tens of anti-atoms.  Second, the deflection methodology requires that the electric field everywhere point in the same direction over the entire length of the trap.  The stochastic acceleration methodology has no such requirement, and the potential can be inverted over a short distance, resulting in much larger electric fields. For instance, if the potential oscillated between that shown in Fig.~\ref{Fig1SchematicFields}d, and its inverse, the average change in the field would be approximately $10\,\mathrm{V}/\mathrm{mm}$, a 20-fold increase over the fields obtained from the potential in Fig.~\ref{Fig1SchematicFields}c. Third, and most important, we can improve the sensitivity of the measurement by taking advantage of the $\sqrt{N}$ scaling of Eq.~(\ref{StochasticLimit}) on the number of acceleration cycles.  As anti-atoms can be held for a very long time \cite{andr:11a}, $N$ can be very large.  We note, however, that this methodology does not yield the sign of any observed putative charge.

\subsection{Numerical Determination of $Q$}

While we can analytically estimate the critical $Q$ from Eq.~\ref{StochasticLimit}, our estimate would rely on a parameter, $\Deltapot$, which can only be approximated analytically.  Furthermore, the random walk in energy-space model on which Eq.~\ref{StochasticLimit} is based has faults, most obviously that it would allow energies to become negative.  Thus, we use numeric simulations to obtain a more precise estimate of the critical $Q$.  These simulations model the anti-atom equation of motion,
\begin{equation}
M\frac{d^2\boldsymbol{\pos}}{d t^2}=\nabla [\muHbarb\cdot\mathbf{B}(\boldsymbol{\pos},t)]
+ Qe[\mathbf{E}(\boldsymbol{\pos},t)+\dot{\boldsymbol{\pos}}\times\mathbf{B}(\boldsymbol{\pos},t)],
\label{EQM}
\end{equation}
where  $\boldsymbol{\pos}$ is the center of mass position of the anti-atom, and $E(\boldsymbol{\pos},t)$ and $B(\boldsymbol{\pos},t)$ are the position and time dependent electric and magnetic fields. For the low-field-seeking anti-atoms modeled here, the magnetic moment $\muHbarb$ and $\mathbf{B}$ are anti-aligned.  Detailed descriptions of similar simulations and various benchmarking tests have been given in prior publications \cite{amol:12, amol:13,amol:14a}.

\begin{figure}[h]
\centerline{\includegraphics[width=3.4in]{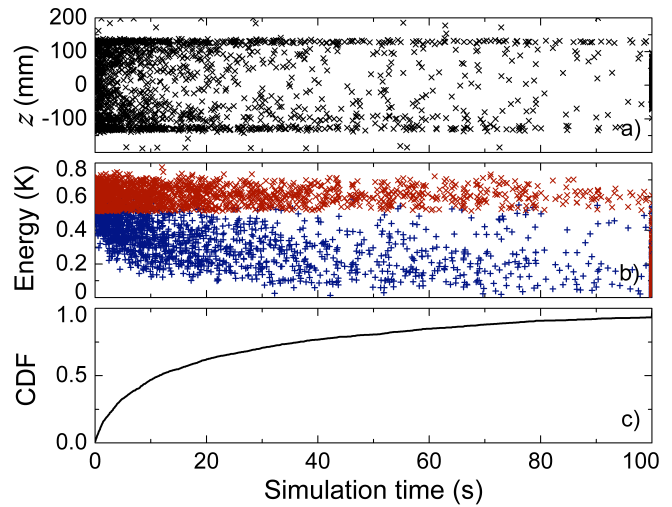}}
\caption{{\bf Typical simulation results} Stochastic acceleration simulation results for $Q=5\times 10^{-10}$, $\DeltaPhi=\pm 350\,$V, $\tstep=0.3\,$ms, and $\tsigma/\tstep=0.2$. (a) The axial location $z$ of the annihilations shows that anti-atoms tend to escape near the octupole ends; particles shown at $t=100\,$s correspond to anti-atoms that remain trapped after the end of the acceleration cycles. Plot (b) shows the initial (blue pluses) and final (red x's) energies of the anti-atoms when they annihilate on the trap wall. Plot (c) shows the cumulative distribution function (CDF) of the probability of escape of the anti-atoms.}
\label{Simulation}
\end{figure}

The results of a typical simulation are shown in Fig~\ref{Simulation}.  The simulation ran for $\THeat=100\,$s for anti-atoms with a putative charge of $Q=5\times 10^{-10}$, and with the $\DeltaPhi=\pm 350\,$V accelerating potentials shown in Fig.~\ref{Fig1SchematicFields}d.  On average, the potentials were inverted every $\tstep=0.3\,$ms, but these switching times were randomized with a standard deviation of $\tsigma=60\,\mu$s.  (The reason for the randomization, which follows a uniform distribution, will be discussed later.)  During the $100\,$s that the simulation ran, approximately $93$\% of the anti-atoms were forced out of the trap and annihilated on the trap wall. Figure~\ref{Simulation}a shows the $z$ locations of these annihilations.  Most of the anti-atoms annihilated near the axial ends of the trap; either on the axial step present at the end, or in the potential well ``holes'' (see Fig.~4 of Ref.~\cite{faja:04} or Ref~\cite{bert:05}) created by the interaction between the octupole and mirror coils. As expected, Fig.~\ref{Simulation}b shows that deeply trapped anti-atoms (anti-atoms with relatively little initial kinetic energy) take the longest to acquire enough energy to escape the trap.  Figure~\ref{Simulation}c plots the cumulative escaped fraction as a function of time.  At about $\thalf=11.6\,$s, half the anti-atoms have escaped.  As the number of simulated anti-atoms is large ($\sim 3000$), the error in determining $\thalf$ is not large; using Greenwood's formula \cite{lawl:03,empi:13}, we can calculate the interval, $10.4\,{\rm s}<\thalf<12.8\,{\rm s}$, where the true $\thalf$ can be found with at least 95\% confidence.

\begin{figure}[h]
\centerline{\includegraphics[width=3.3in]{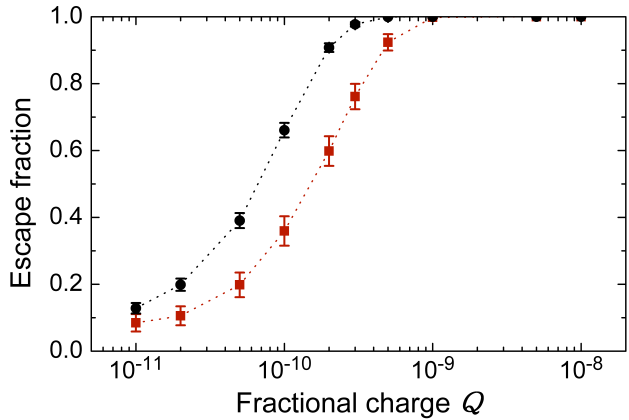}}
\caption{{\bf Escaped Fraction} Fraction of anti-atoms with assumed charge Q that have escaped for $\THeat=500\,$s (black circles) and $\THeat=100\,$s (dark red squares.)  The error bars are determined from the CDF bounds in Fig.~\ref{Simulation}c and establish a 95\% confidence interval for the true escape probability. The other simulation parameters were $\DeltaPhi=\pm 350\,$V, $\tstep=0.3\,$ms, and $\tsigma/\tstep=0.2$.}
\label{EscapeFract}
\end{figure}

\begin{figure}[h]
\centerline{\includegraphics[width=3.3in]{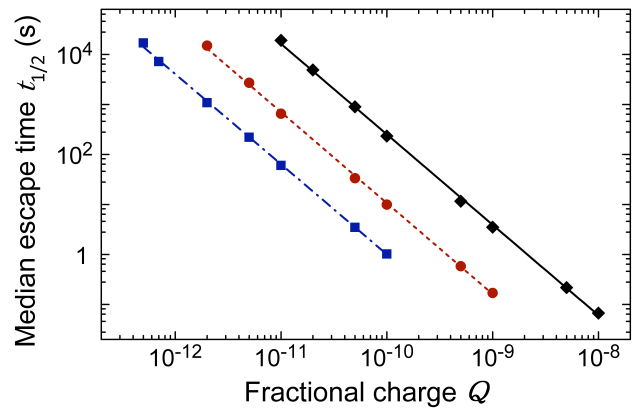}}
\caption{{\bf Stochastic scaling} The acceleration time $\thalf$ for which half the anti-atoms would be expelled as a function of $Q$, as found by simulation. Simulations are shown for our standard trap conditions (black solid lines and diamonds; $\Depth=540\,$mK, $\tstep=0.3\,$ms), with laser cooling (red dashed lines and circles; $\Depth=30\,$mK, $\tstep=1.3\,$ms and an anti-atom temperature $T=20\,$mK), and with laser and adiabatic cooling (blue dot-dashed lines and squares; $\Depth=3\,$mK, $\tstep=4\,$ms and an anti-atom temperature $T=2\,$mK). In all cases, the other simulation parameters were $\DeltaPhi=\pm 350\,$V, and $\tsigma/\tstep=0.2$, and the statistical error at each point is smaller than the point symbol.  Also shown are lines scaling as $Q\propto (\thalf)^{-0.56}$, which describes the relationships between $Q$ and $\thalf$.  The critical $Q$ can be found be setting $\thalf$ to the total acceleration time $\THeat$, and then finding the corresponding $Q$.}
\label{StochasticScaling}
\end{figure}

As shown in Fig.~\ref{EscapeFract}, a large number of simulations similar to those in Fig~\ref{Simulation} can be compiled into a parameter scan showing the fraction of anti-atoms that have escaped as a function of their charge $Q$ for a fixed $\THeat$.  Alternately, a simulation parameter survey can be used to find $\thalf$ as a function of $Q$, and the results compiled into Fig.~\ref{StochasticScaling}.

From Fig.~\ref{StochasticScaling}, we find numerically that $Q$ and $\thalf$ scale as $Q\propto (\thalf)^{-0.56}$.  This differs slightly from the diffusive prediction of Eq.~\ref{StochasticLimit}, which suggests $Q\propto (\thalf)^{-0.50}$.  Part of the difference between these two scaling relations stems from the hard upper energy cutoff assumed in Eq.~\ref{StochasticLimit}; in the analytic calculation, anti-atoms are assumed to annihilate immediately on reaching the magnetic well depth $\Depth$.  The simulations, however, show that there exist quasi-stable orbits with total energy greater than $\Depth$. Anti-atoms on these orbits will remain trapped for some time \cite{coak:05,andr:11a, amol:12}.  Evidence for these ``quasitrapped'' anti-atoms can be seen in Fig.~\ref{Simulation}b, which shows that the final energies of many simulated anti-atoms are well above the trap depth of $\Depth=540\,$mK.  If the simulations are prematurely halted when the anti-atoms' energies exceed $\Depth$ rather than when they annihilate on the trap wall, the simulation scaling changes to $Q\propto (\thalf)^{-0.52}$, significantly closer to the analytic prediction \cite{baqu:13}.  The remaining difference may be due to the lack of a zero bound in the analytic calculation, and the existence of quasi-periodic orbits in the simulation.

\subsection{Comparison of Stochastic Acceleration with Resonant and Autoresonant Acceleration}

\begin{figure}[h]
\centerline{\includegraphics[width=3.3in]{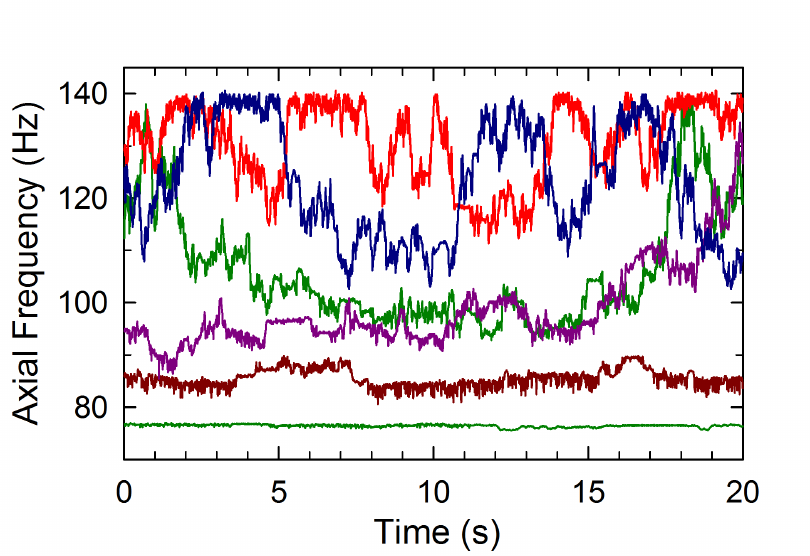}}
\caption{{\bf Axial oscillation frequency} Typical axial oscillation frequencies for undriven anti-atoms.  Note that while the frequency generally appears chaotic, there are periods of stability.  This is particularly true for the frequency of low energy anti-atoms such as is shown by the green, bottom curve in this graph. }
\label{AxialFreq}
\end{figure}

Since our stochastic acceleration scales with the square root of the number of acceleration cycles, it is not as efficient as resonant or autoresonant (swept frequency) acceleration \cite{faja:01a}.  Unfortunately we cannot use either of the latter approaches. Resonant acceleration, i.e.\ simply driving the anti-atoms with an electric field that oscillates at their predicted bounce frequency, requires that a single frequency be resonant with all anti-atoms.   While the anti-atoms do undergo an approximately sinusoidal oscillation in $z$ (see Fig.~C1a in Ref.~\citenum{amol:12}), the frequency of this oscillation is not unique.  Numerical simulations show that the oscillation frequency increases with energy; the system is ``stiff.'' This would appear to make the system a candidate for an autoresonant  drive.  However, autoresonance is best at capturing and exciting initially stationary particles \cite{faja:99d}.  In our case, the anti-atoms are already excited and few would be captured.  Moreover, because of exchanges between parallel and perpendicular energy, the axial oscillation frequency exhibits drifts and shifts (see Fig.~\ref{AxialFreq} for examples in an undriven system).  Any anti-atoms that had been captured in a trapping bucket would quickly escape.  While we have made no formal study, the longitudinal motion generally appears to be chaotic in the long term.

\subsection{Stochasticity}

Stochastic acceleration is essentially a random walk process, and thus requires an element of randomness in the relation between the drive frequency and oscillation frequency.  In many cases, the already-present axial frequency shifts are sufficient. However, Fig.~\ref{AxialFreq} shows that there can be long lasting periods of frequency stability, particularly for low energy anti-atoms, and the anti-atoms may not heat during these periods.  To introduce additional randomness into the system, we modulate $\tstep$ by a random function with uniform distribution and standard deviation $\tsigma$.  Figure~\ref{Sigma-Escape} plots the relation between $\thalf$, the median escape time and $\tsigma$.  It shows that while stochastic acceleration occurs even in the absence of this modulation, the escape time $\tstep$ decreases as the modulation $\tsigma$ increases, reaching a plateau once $\tsigma$ is approximately 20\% of the switching time $\tstep$.  Note that once randomization is introduced in the switching times, it is not necessary to randomize the voltage levels between which the electrode potentials switch. On average, doing so only reduces $\DeltaPhi$, thereby decreasing the energy kicks that the anti-atoms receive, and reducing the acceleration.

\begin{figure}[h]
\centerline{\includegraphics[width=3.3in]{Fig7_sigma.pdf}}
\caption{{\bf Switching time randomization} Median escape time $\thalf$ as a function of normalized drive randomization time $\tsigma/\tstep$, for $Q=1\times 10^{-9}$, $\DeltaPhi=\pm 350\,$V, $\tstep=0.3\,$ms.}
\label{Sigma-Escape}
\end{figure}

\subsection{Switching Time}
To obtain the shortest mean escape times, the switching time $\tstep$ must be optimized to obtain the best bound on $Q$.  In an experiment running for a fixed total time $\THeat$, the number of inversions $N$ is inversely proportional to $\tstep$, thus suggesting a short $\tstep$.  However, if $\tstep$ is too short, anti-atoms will not have time to respond to the electric field before the electric field is inverted, and the size of the energy kick for each inversion will diminish.  The optimal switching time is expected to be comparable to the time it takes an anti-atom to traverse the distance between two oppositely biased electrodes.  For an anti-atom trapped near the top of a $\Depth=540\,$mK well and for the standard electrode configuration, this time is about $0.3\,$ms.  Note that this timescale is much shorter than the orbital periods predicted by Fig.~\ref{AxialFreq}, as here the anti-atoms need only transverse the distance between field reversals, typically between adjacent electrodes, not the entire trap.  (In the experiments to date, discussed in the Introduction, the fields did extend across the entire trap, and the relevant time scale was the orbital bounce period.)

\begin{figure}[h]
\centerline{\includegraphics[width=3.3in]{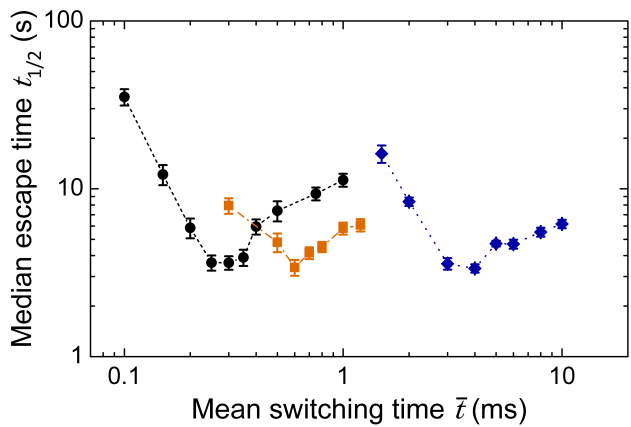}}
\caption{{\bf Switching time optimization} Median escape time $\thalf$ as a function of $\tstep$, for anti-atoms heated with $\DeltaPhi=\pm 350\,$V, and $\tsigma/\tstep=0.2$ for three experimental configurations. The black circles correspond to $Q=10^{-9}$ and to potentials alternating between contiguous electrodes like those shown in Fig.~\ref{Fig1SchematicFields}d by the black line; this is the default configuration generally used elsewhere in this paper. The orange squares correspond to the same $Q$, but with contiguous electrodes joined together and potentials varying between pairs of electrodes, like those shown in Fig.~\ref{Fig1SchematicFields}d by the orange-dot dashed line.  In both these cases, the trap depth is $\Depth=540\,$mK.   The blue diamonds assume colder anti-atoms ($T=2\,$mK) in a shallower trap ($\Depth = 3\,$mK), and $Q=5\times 10^{-11}$.}
\label{Tstep}
\end{figure}

This optimal time is confirmed in Fig.~\ref{Tstep}, which graphs the relation between $\thalf$ and $\tstep$.  The optimal (shortest) $\thalf$ depends on the experimental configuration.
For $\Depth=540\,$mK and the standard electrode configuration, the optimal $\tstep$ is approximately $0.3\,$ms.  As expected, electrically joining adjacent electrodes (see Fig.~\ref{Fig1SchematicFields}d) doubles the optimal $\tstep$. Cooling the anti-atoms by a factor of $100$ increases the optimal $\tstep$ by the expected factor of around $10$.  This is unavoidable, but unfortunate because it decreases the number $N$ of kicks that fit into a fixed acceleration time $\THeat$.  However, the benefits from the lower trapping potential outweigh the disadvantages of the decreased number of kicks, and cooling is, on net, beneficial (see Fig.~\ref{StochasticScaling}).  Fortunately, for a particular configuration, one does not have to hit the optimum exactly.  For $\Depth=540\,$mK and the standard electrode configuration, for example, the escape time $\thalf$ varies by an acceptable factor of $2$ over a mean switching time $\tstep$ range that varies by a factor of around $3$.

\section{Conclusion}
In earlier work, ALPHA established \cite{andr:11a} that antihydrogen can be trapped for at least $1000\,$s; with expected improvements to the vacuum, it is not unreasonable to assume that anti-atoms could be trapped for $10,000\,$s.  This would allow time for $N\approx 10^7$ acceleration cycles; simulations (see Fig.~\ref{StochasticScaling}) then suggest that we could bound $|Q|$ to $3\times 10^{-11}$.  With laser cooled anti-atoms at $20\,$mK \cite{donn:13} we could reduce the trapping potential to perhaps $\Depth=30\,$mK, while still retaining most of the anti-atoms, and the bound would drop to $3\times 10^{-12}$.  Adiabatic expansion cooling of the anti-atoms might reduce their temperature by a further factor of ten, yielding a bound around $ 10^{-12}$.  This bound approaches the limit where antihydrogen polarization effects, studied in Ref.~\citenum{baqu:13}, need to be taken into account.  Other systematic effects are likely to be small, as the relevant experimental parameters (the applied electric potentials and magnetic fields) are well controlled, and, from Eq.~(\ref{StochasticLimit}), expected to enter into the result only linearly.

We note that the conducting tubes in which an anti-atom gravity experiment would take place would exhibit anomalous ``patch'' electric fields \cite{camp:91,amol:14a}, and this could cause a a significant systematic error \cite{witt:68}.  Indeed, a measurement of $Q$ on the order suggested here may be necessary for future precision gravity measurements \cite{kell:08,char:11a,zhmo:13,hami:14}.

We also note that the charge anomaly of the antiproton, $\big||q_{\pbar}|-e\big|/e$, is known \cite{beri:12,hori:11,gabr:99a} to be less than $7\times 10^{-10}$ by measurements \cite{hori:06} on $\pbar{\mathrm{He}}^{+}$, while the charge anomaly of the positron \cite{fee:93,beri:12} is less well known: $|(q_{e^+}-e)/e|< 2.5 \times 10^{-8}$, determined by measurements of the positron cyclotron frequency and the positronium Rydberg constant \cite{hugh:92}.  Thus, under the assumption that the positron and antiproton charges add exactly to form the charge of the antihydrogen atom, an improved measurement of the antihydrogen charge would improve the bound on the positron charge anomaly.

\section{Acknowledgements}
This work was supported by the DOE, NSF, LBNL-LDRD (USA); the experimental data was acquired by the ALPHA collaboration with additional support from: CNPq, FINEP/RENAFAE (Brazil); ISF (Israel); FNU (Denmark); VR (Sweden); NSERC, NRC/TRIUMF, AITF, FQRNT (Canada); and EPSRC, the Royal Society and the Leverhulme Trust (UK). The ALPHA collaboration is grateful for the efforts of the CERN AD team, without which the experimental data could not have been collected. This article constitutes part of the Ph.D work of MB-R.


\begin{thebibliography}{40}
\expandafter\ifx\csname natexlab\endcsname\relax\def\natexlab#1{#1}\fi
\expandafter\ifx\csname bibnamefont\endcsname\relax
  \def\bibnamefont#1{#1}\fi
\expandafter\ifx\csname bibfnamefont\endcsname\relax
  \def\bibfnamefont#1{#1}\fi
\expandafter\ifx\csname citenamefont\endcsname\relax
  \def\citenamefont#1{#1}\fi
\expandafter\ifx\csname url\endcsname\relax
  \def\url#1{\texttt{#1}}\fi
\expandafter\ifx\csname urlprefix\endcsname\relax\def\urlprefix{URL }\fi
\providecommand{\bibinfo}[2]{#2}
\providecommand{\eprint}[2][]{\url{#2}}

\bibitem[{\citenamefont{Andresen et~al.}(2010)\citenamefont{Andresen,
  Ashkezari, Baquero-Ruiz, Bertsche, Bowe, Butler, Cesar, Chapman, Charlton,
  Deller et~al.}}]{andr:10a}
\bibinfo{author}{\bibfnamefont{G.~B.} \bibnamefont{Andresen}},
  \bibinfo{author}{\bibfnamefont{M.~D.} \bibnamefont{Ashkezari}},
  \bibinfo{author}{\bibfnamefont{M.}~\bibnamefont{Baquero-Ruiz}},
  \bibinfo{author}{\bibfnamefont{W.}~\bibnamefont{Bertsche}},
  \bibinfo{author}{\bibfnamefont{P.~D.} \bibnamefont{Bowe}},
  \bibinfo{author}{\bibfnamefont{E.}~\bibnamefont{Butler}},
  \bibinfo{author}{\bibfnamefont{C.~L.} \bibnamefont{Cesar}},
  \bibinfo{author}{\bibfnamefont{S.}~\bibnamefont{Chapman}},
  \bibinfo{author}{\bibfnamefont{M.}~\bibnamefont{Charlton}},
  \bibinfo{author}{\bibfnamefont{A.}~\bibnamefont{Deller}},
  \bibnamefont{et~al.} (\bibinfo{collaboration}{ALPHA Collaboration}),
  \bibinfo{journal}{Nature} \textbf{\bibinfo{volume}{468}},
  \bibinfo{pages}{673} (\bibinfo{year}{2010}).

\bibitem[{\citenamefont{Andresen
  et~al.}(2011{\natexlab{a}})\citenamefont{Andresen, Ashkezari, Baquero-Ruiz,
  Bertsche, Bowe, Butler, Cesar, Charlton, Deller, Eriksson et~al.}}]{andr:11a}
\bibinfo{author}{\bibfnamefont{G.~B.} \bibnamefont{Andresen}},
  \bibinfo{author}{\bibfnamefont{M.~D.} \bibnamefont{Ashkezari}},
  \bibinfo{author}{\bibfnamefont{M.}~\bibnamefont{Baquero-Ruiz}},
  \bibinfo{author}{\bibfnamefont{W.}~\bibnamefont{Bertsche}},
  \bibinfo{author}{\bibfnamefont{P.~D.} \bibnamefont{Bowe}},
  \bibinfo{author}{\bibfnamefont{E.}~\bibnamefont{Butler}},
  \bibinfo{author}{\bibfnamefont{C.~L.} \bibnamefont{Cesar}},
  \bibinfo{author}{\bibfnamefont{M.}~\bibnamefont{Charlton}},
  \bibinfo{author}{\bibfnamefont{A.}~\bibnamefont{Deller}},
  \bibinfo{author}{\bibfnamefont{S.}~\bibnamefont{Eriksson}},
  \bibnamefont{et~al.} (\bibinfo{collaboration}{ALPHA Collaboration}),
  \bibinfo{journal}{Nature Phys.} \textbf{\bibinfo{volume}{7}},
  \bibinfo{pages}{558} (\bibinfo{year}{2011}{\natexlab{a}}).

\bibitem[{\citenamefont{Andresen
  et~al.}(2011{\natexlab{b}})\citenamefont{Andresen, Ashkezari, Baquero-Ruiz,
  Bertsche, Bowe, Bray, Butler, Cesar, Chapman, Charlton et~al.}}]{andr:11b}
\bibinfo{author}{\bibfnamefont{G.~B.} \bibnamefont{Andresen}},
  \bibinfo{author}{\bibfnamefont{M.~D.} \bibnamefont{Ashkezari}},
  \bibinfo{author}{\bibfnamefont{M.}~\bibnamefont{Baquero-Ruiz}},
  \bibinfo{author}{\bibfnamefont{W.}~\bibnamefont{Bertsche}},
  \bibinfo{author}{\bibfnamefont{P.~D.} \bibnamefont{Bowe}},
  \bibinfo{author}{\bibfnamefont{C.~C.} \bibnamefont{Bray}},
  \bibinfo{author}{\bibfnamefont{E.}~\bibnamefont{Butler}},
  \bibinfo{author}{\bibfnamefont{C.~L.} \bibnamefont{Cesar}},
  \bibinfo{author}{\bibfnamefont{S.}~\bibnamefont{Chapman}},
  \bibinfo{author}{\bibfnamefont{M.}~\bibnamefont{Charlton}},
  \bibnamefont{et~al.} (\bibinfo{collaboration}{ALPHA Collaboration}),
  \bibinfo{journal}{Phys. Lett. B} \textbf{\bibinfo{volume}{695}},
  \bibinfo{pages}{95} (\bibinfo{year}{2011}{\natexlab{b}}).

\bibitem[{\citenamefont{Gabrielse et~al.}(2012)\citenamefont{Gabrielse, Kalra,
  Kolthammer, McConnell, Richerme, Grzonka, Oelert, Sefzick, Zielinski,
  Fitzakerley et~al.}}]{gabr:12}
\bibinfo{author}{\bibfnamefont{G.}~\bibnamefont{Gabrielse}},
  \bibinfo{author}{\bibfnamefont{R.}~\bibnamefont{Kalra}},
  \bibinfo{author}{\bibfnamefont{W.~S.} \bibnamefont{Kolthammer}},
  \bibinfo{author}{\bibfnamefont{R.}~\bibnamefont{McConnell}},
  \bibinfo{author}{\bibfnamefont{P.}~\bibnamefont{Richerme}},
  \bibinfo{author}{\bibfnamefont{D.}~\bibnamefont{Grzonka}},
  \bibinfo{author}{\bibfnamefont{W.}~\bibnamefont{Oelert}},
  \bibinfo{author}{\bibfnamefont{T.}~\bibnamefont{Sefzick}},
  \bibinfo{author}{\bibfnamefont{M.}~\bibnamefont{Zielinski}},
  \bibinfo{author}{\bibfnamefont{D.~W.} \bibnamefont{Fitzakerley}},
  \bibnamefont{et~al.} (\bibinfo{collaboration}{ATRAP Collaboration}),
  \bibinfo{journal}{Phys. Rev. Lett.} \textbf{\bibinfo{volume}{108}},
  \bibinfo{pages}{113002} (\bibinfo{year}{2012}).

\bibitem[{\citenamefont{Amole et~al.}(2012{\natexlab{a}})\citenamefont{Amole,
  Ashkezari, Baquero-Ruiz, Bertsche, Bowe, Butler, Capra, Cesar, Charlton,
  Deller et~al.}}]{amol:12a}
\bibinfo{author}{\bibfnamefont{C.}~\bibnamefont{Amole}},
  \bibinfo{author}{\bibfnamefont{M.~D.} \bibnamefont{Ashkezari}},
  \bibinfo{author}{\bibfnamefont{M.}~\bibnamefont{Baquero-Ruiz}},
  \bibinfo{author}{\bibfnamefont{W.}~\bibnamefont{Bertsche}},
  \bibinfo{author}{\bibfnamefont{P.~D.} \bibnamefont{Bowe}},
  \bibinfo{author}{\bibfnamefont{E.}~\bibnamefont{Butler}},
  \bibinfo{author}{\bibfnamefont{A.}~\bibnamefont{Capra}},
  \bibinfo{author}{\bibfnamefont{C.~L.} \bibnamefont{Cesar}},
  \bibinfo{author}{\bibfnamefont{M.}~\bibnamefont{Charlton}},
  \bibinfo{author}{\bibfnamefont{A.}~\bibnamefont{Deller}},
  \bibnamefont{et~al.} (\bibinfo{collaboration}{ALPHA Collaboration}),
  \bibinfo{journal}{Nature} \textbf{\bibinfo{volume}{483}},
  \bibinfo{pages}{439} (\bibinfo{year}{2012}{\natexlab{a}}).

\bibitem[{\citenamefont{Amole et~al.}(2013)\citenamefont{Amole, Ashkezari,
  Baquero-Ruiz, Bertsche, Butler, Capra, Cesar, Charlton, Eriksson, Fajans
  et~al.}}]{amol:13}
\bibinfo{author}{\bibfnamefont{C.}~\bibnamefont{Amole}},
  \bibinfo{author}{\bibfnamefont{M.~D.} \bibnamefont{Ashkezari}},
  \bibinfo{author}{\bibfnamefont{M.}~\bibnamefont{Baquero-Ruiz}},
  \bibinfo{author}{\bibfnamefont{W.}~\bibnamefont{Bertsche}},
  \bibinfo{author}{\bibfnamefont{E.}~\bibnamefont{Butler}},
  \bibinfo{author}{\bibfnamefont{A.}~\bibnamefont{Capra}},
  \bibinfo{author}{\bibfnamefont{C.~L.} \bibnamefont{Cesar}},
  \bibinfo{author}{\bibfnamefont{M.}~\bibnamefont{Charlton}},
  \bibinfo{author}{\bibfnamefont{S.}~\bibnamefont{Eriksson}},
  \bibinfo{author}{\bibfnamefont{J.}~\bibnamefont{Fajans}},
  \bibnamefont{et~al.} (\bibinfo{collaboration}{ALPHA Collaboration}),
  \bibinfo{journal}{Nat. Commun.} \textbf{\bibinfo{volume}{4}},
  \bibinfo{pages}{1785} (\bibinfo{year}{2013}).

\bibitem[{\citenamefont{Kellerbauer et~al.}(2008)\citenamefont{Kellerbauer,
  Amoretti, Belov, Bonomi, Boscolo, Brusa, B\"{u}chner, Byakov, Cabaret, Canali
  et~al.}}]{kell:08}
\bibinfo{author}{\bibfnamefont{A.}~\bibnamefont{Kellerbauer}},
  \bibinfo{author}{\bibfnamefont{M.}~\bibnamefont{Amoretti}},
  \bibinfo{author}{\bibfnamefont{A.~S.} \bibnamefont{Belov}},
  \bibinfo{author}{\bibfnamefont{G.}~\bibnamefont{Bonomi}},
  \bibinfo{author}{\bibfnamefont{I.}~\bibnamefont{Boscolo}},
  \bibinfo{author}{\bibfnamefont{R.~S.} \bibnamefont{Brusa}},
  \bibinfo{author}{\bibfnamefont{M.}~\bibnamefont{B\"{u}chner}},
  \bibinfo{author}{\bibfnamefont{V.~M.} \bibnamefont{Byakov}},
  \bibinfo{author}{\bibfnamefont{L.}~\bibnamefont{Cabaret}},
  \bibinfo{author}{\bibfnamefont{C.}~\bibnamefont{Canali}},
  \bibnamefont{et~al.}, \bibinfo{journal}{Nucl. Instrum. Meth. Phys. Res. B}
  \textbf{\bibinfo{volume}{266}}, \bibinfo{pages}{351} (\bibinfo{year}{2008}).

\bibitem[{cha()}]{char:11a}
\bibinfo{note}{G. Chardin, {\em et al.}, Proposal to measure the gravitational
  behaviour of antihydrogen at rest, Report No. CERN-SPSC-2011-029 / SPSC-P-342
  30/09/2011 (CERN, 2011)}.

\bibitem[{\citenamefont{Ashkezari et~al.}(2012)\citenamefont{Ashkezari,
  Andresen, Baquero-Ruiz, Bertsche, Bowe, Butler, Cesar, Chapman, Charlton,
  Deller et~al.}}]{ashk:12}
\bibinfo{author}{\bibfnamefont{M.~D.} \bibnamefont{Ashkezari}},
  \bibinfo{author}{\bibfnamefont{G.~B.} \bibnamefont{Andresen}},
  \bibinfo{author}{\bibfnamefont{M.}~\bibnamefont{Baquero-Ruiz}},
  \bibinfo{author}{\bibfnamefont{W.}~\bibnamefont{Bertsche}},
  \bibinfo{author}{\bibfnamefont{P.~D.} \bibnamefont{Bowe}},
  \bibinfo{author}{\bibfnamefont{E.}~\bibnamefont{Butler}},
  \bibinfo{author}{\bibfnamefont{C.~L.} \bibnamefont{Cesar}},
  \bibinfo{author}{\bibfnamefont{S.}~\bibnamefont{Chapman}},
  \bibinfo{author}{\bibfnamefont{M.}~\bibnamefont{Charlton}},
  \bibinfo{author}{\bibfnamefont{A.}~\bibnamefont{Deller}},
  \bibnamefont{et~al.} (\bibinfo{collaboration}{ALPHA Collaboration}),
  \bibinfo{journal}{Hyperfine Interact.} \textbf{\bibinfo{volume}{212}},
  \bibinfo{pages}{81} (\bibinfo{year}{2012}).

\bibitem[{\citenamefont{Zhmoginov et~al.}(2013)\citenamefont{Zhmoginov,
  Charman, Shalloo, Fajans, and Wurtele}}]{zhmo:13}
\bibinfo{author}{\bibfnamefont{A.~I.} \bibnamefont{Zhmoginov}},
  \bibinfo{author}{\bibfnamefont{A.~E.} \bibnamefont{Charman}},
  \bibinfo{author}{\bibfnamefont{R.}~\bibnamefont{Shalloo}},
  \bibinfo{author}{\bibfnamefont{J.}~\bibnamefont{Fajans}}, \bibnamefont{and}
  \bibinfo{author}{\bibfnamefont{J.~S.} \bibnamefont{Wurtele}},
  \bibinfo{journal}{Class. and Quantum Grav.} \textbf{\bibinfo{volume}{30}},
  \bibinfo{pages}{205014} (\bibinfo{year}{2013}).

\bibitem[{\citenamefont{Cesar et~al.}(2009)\citenamefont{Cesar, Andresen,
  Bertsche, Bowe, Bray, Butler, Chapman, Charlton, Fajans, Fujiwara
  et~al.}}]{cesa:09a}
\bibinfo{author}{\bibfnamefont{C.~L.} \bibnamefont{Cesar}},
  \bibinfo{author}{\bibfnamefont{G.~B.} \bibnamefont{Andresen}},
  \bibinfo{author}{\bibfnamefont{W.}~\bibnamefont{Bertsche}},
  \bibinfo{author}{\bibfnamefont{P.~D.} \bibnamefont{Bowe}},
  \bibinfo{author}{\bibfnamefont{C.~C.} \bibnamefont{Bray}},
  \bibinfo{author}{\bibfnamefont{E.}~\bibnamefont{Butler}},
  \bibinfo{author}{\bibfnamefont{S.}~\bibnamefont{Chapman}},
  \bibinfo{author}{\bibfnamefont{M.}~\bibnamefont{Charlton}},
  \bibinfo{author}{\bibfnamefont{J.}~\bibnamefont{Fajans}},
  \bibinfo{author}{\bibfnamefont{M.~C.} \bibnamefont{Fujiwara}},
  \bibnamefont{et~al.}, \bibinfo{journal}{Can.\ J.\ of Phys.}
  \textbf{\bibinfo{volume}{87}}, \bibinfo{pages}{791} (\bibinfo{year}{2009}).

\bibitem[{\citenamefont{Hamilton et~al.}(2014)\citenamefont{Hamilton,
  Zhmoginov, Robicheaux, Fajans, Wurtele, and M\"uller}}]{hami:14}
\bibinfo{author}{\bibfnamefont{P.}~\bibnamefont{Hamilton}},
  \bibinfo{author}{\bibfnamefont{A.}~\bibnamefont{Zhmoginov}},
  \bibinfo{author}{\bibfnamefont{F.}~\bibnamefont{Robicheaux}},
  \bibinfo{author}{\bibfnamefont{J.}~\bibnamefont{Fajans}},
  \bibinfo{author}{\bibfnamefont{J.~S.} \bibnamefont{Wurtele}},
  \bibnamefont{and} \bibinfo{author}{\bibfnamefont{H.}~\bibnamefont{M\"uller}},
  \bibinfo{journal}{Phys. Rev. Lett.} \textbf{\bibinfo{volume}{112}},
  \bibinfo{pages}{121102} (\bibinfo{year}{2014}).

\bibitem[{\citenamefont{Bressi et~al.}(2011)\citenamefont{Bressi, Carugno,
  Della~Valle, Galeazzi, Ruoso, and Sartori}}]{bres:11}
\bibinfo{author}{\bibfnamefont{G.}~\bibnamefont{Bressi}},
  \bibinfo{author}{\bibfnamefont{G.}~\bibnamefont{Carugno}},
  \bibinfo{author}{\bibfnamefont{F.}~\bibnamefont{Della~Valle}},
  \bibinfo{author}{\bibfnamefont{G.}~\bibnamefont{Galeazzi}},
  \bibinfo{author}{\bibfnamefont{G.}~\bibnamefont{Ruoso}}, \bibnamefont{and}
  \bibinfo{author}{\bibfnamefont{G.}~\bibnamefont{Sartori}},
  \bibinfo{journal}{Phys. Rev. A} \textbf{\bibinfo{volume}{83}},
  \bibinfo{pages}{052101} (\bibinfo{year}{2011}).

\bibitem[{\citenamefont{Quigg}(1997)}]{quig:97}
\bibinfo{author}{\bibfnamefont{C.}~\bibnamefont{Quigg}},
  \emph{\bibinfo{title}{Gauge theories of the strong, weak, and electromagnetic
  interactions}} (\bibinfo{publisher}{Westview Press}, \bibinfo{year}{1997}).

\bibitem[{\citenamefont{Amole et~al.}(2014)\citenamefont{Amole, Ashkezari,
  Baquero-Ruiz, Bertsche, Butler, Capra, Cesar, Charlton, Eriksson, Fajans
  et~al.}}]{amol:14a}
\bibinfo{author}{\bibfnamefont{C.}~\bibnamefont{Amole}},
  \bibinfo{author}{\bibfnamefont{M.~D.} \bibnamefont{Ashkezari}},
  \bibinfo{author}{\bibfnamefont{M.}~\bibnamefont{Baquero-Ruiz}},
  \bibinfo{author}{\bibfnamefont{W.}~\bibnamefont{Bertsche}},
  \bibinfo{author}{\bibfnamefont{E.}~\bibnamefont{Butler}},
  \bibinfo{author}{\bibfnamefont{A.}~\bibnamefont{Capra}},
  \bibinfo{author}{\bibfnamefont{C.~L.} \bibnamefont{Cesar}},
  \bibinfo{author}{\bibfnamefont{M.}~\bibnamefont{Charlton}},
  \bibinfo{author}{\bibfnamefont{S.}~\bibnamefont{Eriksson}},
  \bibinfo{author}{\bibfnamefont{J.}~\bibnamefont{Fajans}},
  \bibnamefont{et~al.}, \bibinfo{journal}{Nat. Commun.}
  \textbf{\bibinfo{volume}{5}}, \bibinfo{pages}{3955} (\bibinfo{year}{2014}).

\bibitem[{\citenamefont{Sturrock}(1966)}]{stur:66}
\bibinfo{author}{\bibfnamefont{P.~A.} \bibnamefont{Sturrock}},
  \bibinfo{journal}{Phys. Rev.} \textbf{\bibinfo{volume}{141}},
  \bibinfo{pages}{186} (\bibinfo{year}{1966}).

\bibitem[{\citenamefont{Tsushima and Ishihara}(2009)}]{tsus:09}
\bibinfo{author}{\bibfnamefont{A.}~\bibnamefont{Tsushima}} \bibnamefont{and}
  \bibinfo{author}{\bibfnamefont{O.}~\bibnamefont{Ishihara}},
  \bibinfo{journal}{J. Plasma Fusion Res.} \textbf{\bibinfo{volume}{8}},
  \bibinfo{pages}{65} (\bibinfo{year}{2009}).

\bibitem[{\citenamefont{Pritchard}(1983)}]{prit:83}
\bibinfo{author}{\bibfnamefont{D.~E.} \bibnamefont{Pritchard}},
  \bibinfo{journal}{Phys.\ Rev.\ Lett.} \textbf{\bibinfo{volume}{51}},
  \bibinfo{pages}{1336} (\bibinfo{year}{1983}).

\bibitem[{\citenamefont{Bertsche et~al.}(2006)\citenamefont{Bertsche, Boston,
  Bowe, Cesar, Chapman, Charlton, Chartier, Deutsch, Fajans, Fujiwara
  et~al.}}]{bert:06}
\bibinfo{author}{\bibfnamefont{W.}~\bibnamefont{Bertsche}},
  \bibinfo{author}{\bibfnamefont{A.}~\bibnamefont{Boston}},
  \bibinfo{author}{\bibfnamefont{P.~D.} \bibnamefont{Bowe}},
  \bibinfo{author}{\bibfnamefont{C.~L.} \bibnamefont{Cesar}},
  \bibinfo{author}{\bibfnamefont{S.}~\bibnamefont{Chapman}},
  \bibinfo{author}{\bibfnamefont{M.}~\bibnamefont{Charlton}},
  \bibinfo{author}{\bibfnamefont{M.}~\bibnamefont{Chartier}},
  \bibinfo{author}{\bibfnamefont{A.}~\bibnamefont{Deutsch}},
  \bibinfo{author}{\bibfnamefont{J.}~\bibnamefont{Fajans}},
  \bibinfo{author}{\bibfnamefont{M.~C.} \bibnamefont{Fujiwara}},
  \bibnamefont{et~al.}, \bibinfo{journal}{Nucl. Instr. Meth. Phys. Res. A}
  \textbf{\bibinfo{volume}{566}}, \bibinfo{pages}{746} (\bibinfo{year}{2006}).

\bibitem[{\citenamefont{Andresen et~al.}(2008)\citenamefont{Andresen, Bertsche,
  Boston, Bowe, Cesar, Chapman, Charlton, Chartier, Deutsch, Fajans
  et~al.}}]{andr:08b}
\bibinfo{author}{\bibfnamefont{G.~B.} \bibnamefont{Andresen}},
  \bibinfo{author}{\bibfnamefont{W.}~\bibnamefont{Bertsche}},
  \bibinfo{author}{\bibfnamefont{A.}~\bibnamefont{Boston}},
  \bibinfo{author}{\bibfnamefont{P.~D.} \bibnamefont{Bowe}},
  \bibinfo{author}{\bibfnamefont{C.~L.} \bibnamefont{Cesar}},
  \bibinfo{author}{\bibfnamefont{S.}~\bibnamefont{Chapman}},
  \bibinfo{author}{\bibfnamefont{M.}~\bibnamefont{Charlton}},
  \bibinfo{author}{\bibfnamefont{M.}~\bibnamefont{Chartier}},
  \bibinfo{author}{\bibfnamefont{A.}~\bibnamefont{Deutsch}},
  \bibinfo{author}{\bibfnamefont{J.}~\bibnamefont{Fajans}},
  \bibnamefont{et~al.} (\bibinfo{collaboration}{ALPHA Collaboration}),
  \bibinfo{journal}{J. Phys. B: At. Mol. Opt. Phys.}
  \textbf{\bibinfo{volume}{41}}, \bibinfo{pages}{011001}
  (\bibinfo{year}{2008}).

\bibitem[{\citenamefont{Andresen
  et~al.}(2011{\natexlab{c}})\citenamefont{Andresen, Ashkezari, Baquero-Ruiz,
  Bertsche, Bowe, Butler, Carpenter, Cesar, Chapman, Charlton
  et~al.}}]{andr:11d}
\bibinfo{author}{\bibfnamefont{G.~B.} \bibnamefont{Andresen}},
  \bibinfo{author}{\bibfnamefont{M.~D.} \bibnamefont{Ashkezari}},
  \bibinfo{author}{\bibfnamefont{M.}~\bibnamefont{Baquero-Ruiz}},
  \bibinfo{author}{\bibfnamefont{W.}~\bibnamefont{Bertsche}},
  \bibinfo{author}{\bibfnamefont{P.~D.} \bibnamefont{Bowe}},
  \bibinfo{author}{\bibfnamefont{E.}~\bibnamefont{Butler}},
  \bibinfo{author}{\bibfnamefont{P.~T.} \bibnamefont{Carpenter}},
  \bibinfo{author}{\bibfnamefont{C.~L.} \bibnamefont{Cesar}},
  \bibinfo{author}{\bibfnamefont{S.}~\bibnamefont{Chapman}},
  \bibinfo{author}{\bibfnamefont{M.}~\bibnamefont{Charlton}},
  \bibnamefont{et~al.} (\bibinfo{collaboration}{ALPHA Collaboration}),
  \bibinfo{journal}{Phys. Rev. Lett.} \textbf{\bibinfo{volume}{106}},
  \bibinfo{pages}{025002} (\bibinfo{year}{2011}{\natexlab{c}}).

\bibitem[{\citenamefont{Andresen et~al.}(2012)\citenamefont{Andresen,
  Ashkezari, Bertsche, Bowe, Butler, Cesar, Chapman, Charlton, Deller, Eriksson
  et~al.}}]{andr:12}
\bibinfo{author}{\bibfnamefont{G.~B.} \bibnamefont{Andresen}},
  \bibinfo{author}{\bibfnamefont{M.~D.} \bibnamefont{Ashkezari}},
  \bibinfo{author}{\bibfnamefont{W.}~\bibnamefont{Bertsche}},
  \bibinfo{author}{\bibfnamefont{P.~D.} \bibnamefont{Bowe}},
  \bibinfo{author}{\bibfnamefont{E.}~\bibnamefont{Butler}},
  \bibinfo{author}{\bibfnamefont{C.~L.} \bibnamefont{Cesar}},
  \bibinfo{author}{\bibfnamefont{S.}~\bibnamefont{Chapman}},
  \bibinfo{author}{\bibfnamefont{M.}~\bibnamefont{Charlton}},
  \bibinfo{author}{\bibfnamefont{A.}~\bibnamefont{Deller}},
  \bibinfo{author}{\bibfnamefont{S.}~\bibnamefont{Eriksson}},
  \bibnamefont{et~al.} (\bibinfo{collaboration}{ALPHA Collaboration}),
  \bibinfo{journal}{Nucl. Instr. Meth. Phys. Res. A}
  \textbf{\bibinfo{volume}{684}}, \bibinfo{pages}{73} (\bibinfo{year}{2012}).

\bibitem[{\citenamefont{Amole et~al.}(2012{\natexlab{b}})\citenamefont{Amole,
  Andresen, Ashkezari, Baquero-Ruiz, Bertsche, Butler, Cesar, Chapman,
  Charlton, Deller et~al.}}]{amol:12}
\bibinfo{author}{\bibfnamefont{C.}~\bibnamefont{Amole}},
  \bibinfo{author}{\bibfnamefont{G.~B.} \bibnamefont{Andresen}},
  \bibinfo{author}{\bibfnamefont{M.~D.} \bibnamefont{Ashkezari}},
  \bibinfo{author}{\bibfnamefont{M.}~\bibnamefont{Baquero-Ruiz}},
  \bibinfo{author}{\bibfnamefont{W.}~\bibnamefont{Bertsche}},
  \bibinfo{author}{\bibfnamefont{E.}~\bibnamefont{Butler}},
  \bibinfo{author}{\bibfnamefont{C.~L.} \bibnamefont{Cesar}},
  \bibinfo{author}{\bibfnamefont{S.}~\bibnamefont{Chapman}},
  \bibinfo{author}{\bibfnamefont{M.}~\bibnamefont{Charlton}},
  \bibinfo{author}{\bibfnamefont{A.}~\bibnamefont{Deller}},
  \bibnamefont{et~al.} (\bibinfo{collaboration}{ALPHA Collaboration}),
  \bibinfo{journal}{New J. Phys.} \textbf{\bibinfo{volume}{14}},
  \bibinfo{pages}{015010} (\bibinfo{year}{2012}{\natexlab{b}}).

\bibitem[{\citenamefont{Donnan et~al.}(2013)\citenamefont{Donnan, Fujiwara, and
  Robicheaux}}]{donn:13}
\bibinfo{author}{\bibfnamefont{P.~H.} \bibnamefont{Donnan}},
  \bibinfo{author}{\bibfnamefont{M.~C.} \bibnamefont{Fujiwara}},
  \bibnamefont{and}
  \bibinfo{author}{\bibfnamefont{F.}~\bibnamefont{Robicheaux}},
  \bibinfo{journal}{J. Phys. B} \textbf{\bibinfo{volume}{46}},
  \bibinfo{pages}{025302} (\bibinfo{year}{2013}).

\bibitem[{\citenamefont{Baquero-Ruiz}(2013)}]{baqu:13}
\bibinfo{author}{\bibfnamefont{M.}~\bibnamefont{Baquero-Ruiz}}, Ph.D. thesis,
  \bibinfo{school}{U.C. Berkeley} (\bibinfo{year}{2013}).

\bibitem[{\citenamefont{Coakley et~al.}(2005)\citenamefont{Coakley, Doyle,
  Dzhosyuk, Yang, and Huffman}}]{coak:05}
\bibinfo{author}{\bibfnamefont{K.}~\bibnamefont{Coakley}},
  \bibinfo{author}{\bibfnamefont{J.}~\bibnamefont{Doyle}},
  \bibinfo{author}{\bibfnamefont{S.}~\bibnamefont{Dzhosyuk}},
  \bibinfo{author}{\bibfnamefont{L.}~\bibnamefont{Yang}}, \bibnamefont{and}
  \bibinfo{author}{\bibfnamefont{P.}~\bibnamefont{Huffman}},
  \bibinfo{journal}{J. Res. Natl. Stand. Technol.}
  \textbf{\bibinfo{volume}{110}}, \bibinfo{pages}{367} (\bibinfo{year}{2005}).

\bibitem[{\citenamefont{Fajans and Schmidt}(2004)}]{faja:04}
\bibinfo{author}{\bibfnamefont{J.}~\bibnamefont{Fajans}} \bibnamefont{and}
  \bibinfo{author}{\bibfnamefont{A.}~\bibnamefont{Schmidt}},
  \bibinfo{journal}{Nucl. Instr. Meth. Phys. Res. A}
  \textbf{\bibinfo{volume}{521}}, \bibinfo{pages}{318} (\bibinfo{year}{2004}).

\bibitem[{\citenamefont{Bertsche et~al.}(2005)\citenamefont{Bertsche, Boston,
  Bowe, Cesar, Chapman, Charlton, Chartier, Deutsch, Fajans, Fujiwara
  et~al.}}]{bert:05}
\bibinfo{author}{\bibfnamefont{W.}~\bibnamefont{Bertsche}},
  \bibinfo{author}{\bibfnamefont{A.}~\bibnamefont{Boston}},
  \bibinfo{author}{\bibfnamefont{P.}~\bibnamefont{Bowe}},
  \bibinfo{author}{\bibfnamefont{C.}~\bibnamefont{Cesar}},
  \bibinfo{author}{\bibfnamefont{S.}~\bibnamefont{Chapman}},
  \bibinfo{author}{\bibfnamefont{M.}~\bibnamefont{Charlton}},
  \bibinfo{author}{\bibfnamefont{M.}~\bibnamefont{Chartier}},
  \bibinfo{author}{\bibfnamefont{A.}~\bibnamefont{Deutsch}},
  \bibinfo{author}{\bibfnamefont{J.}~\bibnamefont{Fajans}},
  \bibinfo{author}{\bibfnamefont{M.}~\bibnamefont{Fujiwara}},
  \bibnamefont{et~al.}, in \emph{\bibinfo{booktitle}{Low Energy Antiproton
  Physics}}, edited by
  \bibinfo{editor}{\bibfnamefont{D.}~\bibnamefont{Grzonka}},
  \bibinfo{editor}{\bibfnamefont{R.}~\bibnamefont{Czyzykiewicz}},
  \bibinfo{editor}{\bibfnamefont{W.}~\bibnamefont{Oelert}},
  \bibinfo{editor}{\bibfnamefont{T.}~\bibnamefont{Rozek}}, \bibnamefont{and}
  \bibinfo{editor}{\bibfnamefont{P.}~\bibnamefont{Winter}}
  (\bibinfo{publisher}{AIP}, \bibinfo{year}{2005}), vol. \bibinfo{volume}{796},
  p. \bibinfo{pages}{301}.

\bibitem[{\citenamefont{Lawless}(2003)}]{lawl:03}
\bibinfo{author}{\bibfnamefont{J.}~\bibnamefont{Lawless}},
  \emph{\bibinfo{title}{Statistical Models and Methods for Lifetime Data}}
  (\bibinfo{publisher}{John Wiley and Sons, Inc.,}, \bibinfo{address}{Hoboken},
  \bibinfo{year}{2003}), \bibinfo{edition}{2nd} ed.

\bibitem[{emp(2013)}]{empi:13}
\emph{\bibinfo{title}{Empirical cumulative distribution function in {MATLAB}}}
  (\bibinfo{year}{2013}),
  \urlprefix\url{http://www.mathworks.com/help/stats/ecdf.html}.

\bibitem[{\citenamefont{Fajans and Friedland}(2001)}]{faja:01a}
\bibinfo{author}{\bibfnamefont{J.}~\bibnamefont{Fajans}} \bibnamefont{and}
  \bibinfo{author}{\bibfnamefont{L.}~\bibnamefont{Friedland}},
  \bibinfo{journal}{Am.\ J.\ Phys.} \textbf{\bibinfo{volume}{69}},
  \bibinfo{pages}{1096} (\bibinfo{year}{2001}).

\bibitem[{\citenamefont{Fajans et~al.}(1999)\citenamefont{Fajans, Gilson, and
  Friedland}}]{faja:99d}
\bibinfo{author}{\bibfnamefont{J.}~\bibnamefont{Fajans}},
  \bibinfo{author}{\bibfnamefont{E.}~\bibnamefont{Gilson}}, \bibnamefont{and}
  \bibinfo{author}{\bibfnamefont{L.}~\bibnamefont{Friedland}},
  \bibinfo{journal}{Phys.\ Plasmas} \textbf{\bibinfo{volume}{6}},
  \bibinfo{pages}{4497} (\bibinfo{year}{1999}).

\bibitem[{\citenamefont{Camp et~al.}(1991)\citenamefont{Camp, Darling, and
  Brown}}]{camp:91}
\bibinfo{author}{\bibfnamefont{J.~B.} \bibnamefont{Camp}},
  \bibinfo{author}{\bibfnamefont{T.~W.} \bibnamefont{Darling}},
  \bibnamefont{and} \bibinfo{author}{\bibfnamefont{R.~E.} \bibnamefont{Brown}},
  \bibinfo{journal}{J. Appl. Phys.} \textbf{\bibinfo{volume}{69}},
  \bibinfo{pages}{7126} (\bibinfo{year}{1991}).

\bibitem[{\citenamefont{Witteborn and Fairbank}(1968)}]{witt:68}
\bibinfo{author}{\bibfnamefont{F.~C.} \bibnamefont{Witteborn}}
  \bibnamefont{and} \bibinfo{author}{\bibfnamefont{W.~M.}
  \bibnamefont{Fairbank}}, \bibinfo{journal}{Nature}
  \textbf{\bibinfo{volume}{220}}, \bibinfo{pages}{436} (\bibinfo{year}{1968}).

\bibitem[{\citenamefont{Beringer et~al.}(2012)\citenamefont{Beringer, Arguin,
  Barnett, Copic, Dahl, Groom, Lin, Lys, Murayama, Wohl et~al.}}]{beri:12}
\bibinfo{author}{\bibfnamefont{J.}~\bibnamefont{Beringer}},
  \bibinfo{author}{\bibfnamefont{J.~F.} \bibnamefont{Arguin}},
  \bibinfo{author}{\bibfnamefont{R.~M.} \bibnamefont{Barnett}},
  \bibinfo{author}{\bibfnamefont{K.}~\bibnamefont{Copic}},
  \bibinfo{author}{\bibfnamefont{O.}~\bibnamefont{Dahl}},
  \bibinfo{author}{\bibfnamefont{D.~E.} \bibnamefont{Groom}},
  \bibinfo{author}{\bibfnamefont{C.~J.} \bibnamefont{Lin}},
  \bibinfo{author}{\bibfnamefont{J.}~\bibnamefont{Lys}},
  \bibinfo{author}{\bibfnamefont{H.}~\bibnamefont{Murayama}},
  \bibinfo{author}{\bibfnamefont{C.~G.} \bibnamefont{Wohl}},
  \bibnamefont{et~al.} (\bibinfo{collaboration}{Particle Data Group}),
  \bibinfo{journal}{Phys. Rev. D} \textbf{\bibinfo{volume}{86}},
  \bibinfo{pages}{010001} (\bibinfo{year}{2012}).

\bibitem[{\citenamefont{Hori et~al.}(2011)\citenamefont{Hori, S\'{o}t\'{e}r,
  Barna, Dax, Hayano, Friedreich, Juh\'asz, Pask, Widmann, Horv\'ath
  et~al.}}]{hori:11}
\bibinfo{author}{\bibfnamefont{M.}~\bibnamefont{Hori}},
  \bibinfo{author}{\bibfnamefont{A.}~\bibnamefont{S\'{o}t\'{e}r}},
  \bibinfo{author}{\bibfnamefont{D.}~\bibnamefont{Barna}},
  \bibinfo{author}{\bibfnamefont{A.}~\bibnamefont{Dax}},
  \bibinfo{author}{\bibfnamefont{R.}~\bibnamefont{Hayano}},
  \bibinfo{author}{\bibfnamefont{S.}~\bibnamefont{Friedreich}},
  \bibinfo{author}{\bibfnamefont{B.}~\bibnamefont{Juh\'asz}},
  \bibinfo{author}{\bibfnamefont{T.}~\bibnamefont{Pask}},
  \bibinfo{author}{\bibfnamefont{E.}~\bibnamefont{Widmann}},
  \bibinfo{author}{\bibfnamefont{D.}~\bibnamefont{Horv\'ath}},
  \bibnamefont{et~al.}, \bibinfo{journal}{Nature}
  \textbf{\bibinfo{volume}{475}}, \bibinfo{pages}{484} (\bibinfo{year}{2011}).

\bibitem[{\citenamefont{Gabrielse et~al.}(1999)\citenamefont{Gabrielse,
  Khabbaz, Hall, Heimann, Kalinowsky, and Jhe}}]{gabr:99a}
\bibinfo{author}{\bibfnamefont{G.}~\bibnamefont{Gabrielse}},
  \bibinfo{author}{\bibfnamefont{A.}~\bibnamefont{Khabbaz}},
  \bibinfo{author}{\bibfnamefont{D.~S.} \bibnamefont{Hall}},
  \bibinfo{author}{\bibfnamefont{C.}~\bibnamefont{Heimann}},
  \bibinfo{author}{\bibfnamefont{H.}~\bibnamefont{Kalinowsky}},
  \bibnamefont{and} \bibinfo{author}{\bibfnamefont{W.}~\bibnamefont{Jhe}},
  \bibinfo{journal}{Phys. Rev. Lett.} \textbf{\bibinfo{volume}{82}},
  \bibinfo{pages}{3198} (\bibinfo{year}{1999}).

\bibitem[{\citenamefont{Hori et~al.}(2006)\citenamefont{Hori, Dax, Eades,
  Gomikawa, Hayano, Ono, Pirkl, Widmann, Torii, Juh\'asz et~al.}}]{hori:06}
\bibinfo{author}{\bibfnamefont{M.}~\bibnamefont{Hori}},
  \bibinfo{author}{\bibfnamefont{A.}~\bibnamefont{Dax}},
  \bibinfo{author}{\bibfnamefont{J.}~\bibnamefont{Eades}},
  \bibinfo{author}{\bibfnamefont{K.}~\bibnamefont{Gomikawa}},
  \bibinfo{author}{\bibfnamefont{R.~S.} \bibnamefont{Hayano}},
  \bibinfo{author}{\bibfnamefont{N.}~\bibnamefont{Ono}},
  \bibinfo{author}{\bibfnamefont{W.}~\bibnamefont{Pirkl}},
  \bibinfo{author}{\bibfnamefont{E.}~\bibnamefont{Widmann}},
  \bibinfo{author}{\bibfnamefont{H.~A.} \bibnamefont{Torii}},
  \bibinfo{author}{\bibfnamefont{B.}~\bibnamefont{Juh\'asz}},
  \bibnamefont{et~al.}, \bibinfo{journal}{Phys. Rev. Lett.}
  \textbf{\bibinfo{volume}{96}}, \bibinfo{pages}{243401}
  (\bibinfo{year}{2006}).

\bibitem[{\citenamefont{Fee et~al.}(1993)\citenamefont{Fee, Chu, Mills,
  Chichester, Zuckerman, Shaw, and Danzmann}}]{fee:93}
\bibinfo{author}{\bibfnamefont{M.~S.} \bibnamefont{Fee}},
  \bibinfo{author}{\bibfnamefont{S.}~\bibnamefont{Chu}},
  \bibinfo{author}{\bibfnamefont{A.~P.} \bibnamefont{Mills}},
  \bibinfo{author}{\bibfnamefont{R.~J.} \bibnamefont{Chichester}},
  \bibinfo{author}{\bibfnamefont{D.~M.} \bibnamefont{Zuckerman}},
  \bibinfo{author}{\bibfnamefont{E.~D.} \bibnamefont{Shaw}}, \bibnamefont{and}
  \bibinfo{author}{\bibfnamefont{K.}~\bibnamefont{Danzmann}},
  \bibinfo{journal}{Phys. Rev. A} \textbf{\bibinfo{volume}{48}},
  \bibinfo{pages}{192} (\bibinfo{year}{1993}).

\bibitem[{\citenamefont{Hughes and Deutch}(1992)}]{hugh:92}
\bibinfo{author}{\bibfnamefont{R.~J.} \bibnamefont{Hughes}} \bibnamefont{and}
  \bibinfo{author}{\bibfnamefont{B.~I.} \bibnamefont{Deutch}},
  \bibinfo{journal}{Phys. Rev. Lett.} \textbf{\bibinfo{volume}{69}},
  \bibinfo{pages}{578} (\bibinfo{year}{1992}).

\end{thebibliography}


\end{document}